# Structure and Energetics of Graphene Oxide Isomers: Ab Initio Thermodynamic Analysis

Vitaly V. Chaban[1] and Oleg V. Prezhdo[2]

[1] Instituto de Ciência e Tecnologia, Universidade Federal de São Paulo, 12231-280, São José dos Campos, SP, Brazil

[2] Department of Chemistry, University of Southern California, Los Angeles, CA 90089, USA

**Abstract**. Graphene oxide (GO) holds significant promise for electronic devices and nanocomposite materials. A number of models were proposed for GO structure, combining carboxyl, hydroxyl, carbonyl and epoxide groups at different locations. The complexity and variety of GO isomers, whose thermodynamic stability and formation kinetics depend on applied conditions, make determination of GO structure with atomistic precision challenging. We report high level theoretical investigation of multiple molecular configurations, which are anticipated in GO. We conclude that all oxygen containing groups at the GO surface are thermodynamically permitted, whereas the 'edge' positions are systematically more favorable than the 'center' and 'side' positions. We discuss a potentially novel type of chemical bond or bonding reinforcement in GO, which consists of a covalent bond and a strong electrostatic contribution from a polarized graphene plane. We observe and analyze significant modifications of graphene geometry and electronic structure upon oxidation. The reported thermodynamic data guide experiments aimed at deciphering GO chemical composition and structure, and form the basis for predicting GO properties required for nano-technological applications.

TOC Graphic

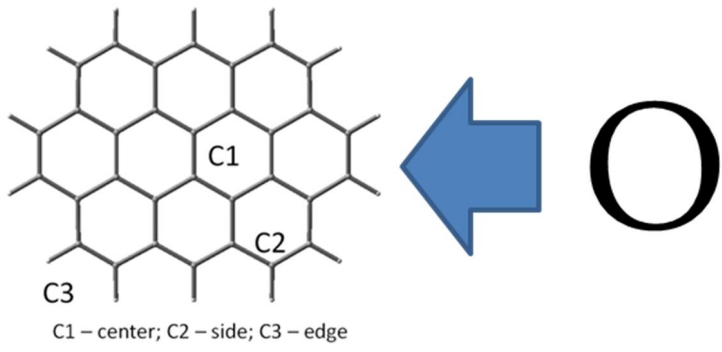

C1 – center; C2 – side; C3 – edge

**Introduction**

Graphene oxide (GO) is a unique material with a high technological promise.[1-29] It is a single monolayer of graphite, i.e. graphene (GR),[2, 3, 13, 26, 30-34] decorated by oxygen containing functional groups (OX),[1, 9] such as hydroxyl (OH), carboxyl (COOH) and epoxide (-O-). A potential of GO for robust applications in optoelectronics, non-volatile memory, biological devices, catalysis, and drug delivery vectors was shown.[25, 35-41] Graphite oxide (graphitic acid) was known long before GR was isolated, characterized and recognized. It could be obtained by oxidation of graphite using a solution of potassium permanganate in sulfuric acid. This approach is still being actively used. It is also a classical method to oxidize many other organic carbonaceous compounds. Graphite oxide may be viewed as a precursor for GR, since oxidation simplifies graphite exfoliation. After exfoliation, a reduction of carbon sheets must be performed to remove oxygen containing groups from the GR surface. Chemical, thermal and electrochemical reduction pathways are possible.

One of the most practical advantages of GO is its straightforward dispersibility in water and numerous organic solvents. Dispersabilities in water typically range from 1 to 4 g L$^{-1}$.[1] In the solubility context, note that content of GO depends on how it was obtained. Therefore, a single well-defined solubility value must not be expected. Subsequent sonication allows high-quality exfoliation from these dispersions. GO can also be dispersed in ceramic and polymer matrices due to the presence of oxygen atoms. GO is an electrical insulator due to the disruption of the honeycomb hexagonal lattice and, therefore, the lack of the sp$^2$ electronic bonding networks present in GR. Once dispersed in the matrix, GO can be reduced to recover electronic properties of GR.[1] A few terms exist in literature for such a reduction product: reduced graphene oxide, chemically reduced graphene oxide, graphene. Reduced GO is often confused with graphene. It should be kept in mind that the reduced GO is neither chemically nor physically identical to pristine graphene. First, reduction of GO is never complete. Second, the reduction reaction significantly modifies carbon lattice, which cannot be restored to the state of pristine GR.

The spatial structure of GO is still not precisely determined despite available elemental compositions. GO exhibits a highly variable stoichiometry. Local arrangement of functional groups is not sufficiently established to provide guidelines for a task-specific synthesis, since not all oxygen containing groups are favorable for applications and further synthetic purposes. The dependence of the degree of oxidation (oxygen content) on synthetic conditions is poorly understood, since GO features many structural defects. The oxidation degree determines the electronic properties of GO. Similar to graphene, GO is a highly electron transparent material thanks to a two-dimensional nature and a low atomic number. Consequently, GO can act as an excellent support film for nanoclusters and macromolecules for investigation using transmission electron microscopy.[41] The mean size of GO sheets is on the order of hundreds nanometers, whereas the mean size of GR sheets can be several microns. Oxidation reactions foster breaking of the graphitic structure into smaller molecular fragments following the same mechanism encountered during vigorous oxidation of hydrocarbons. Many of the first published structural models of GO assume regular lattices with discrete repeat units. The historic model proposed by Hofmann and Holst in 1939 contained multiple epoxide groups, which were uniformly spread across the basal planes of graphite. The simplified molecular formula of the compound according to these authors is $C_2O$. Such a model does not account for certain hydrogen content in GO, as suggested by an elemental analysis. To deal with this 7 years later, Ruess positioned hydroxyl groups in the same plane with epoxide groups. Mermoux accounted for structural similarities between GO and poly(carbon monofluoride).[42] This structure implies complete rehybridization of the $sp^2$ planes in graphite into $sp^3$ cyclohexyl structures.

More recently, Lerf and Klinowski questioned the lattice-based models of GO using the results of nuclear magnetic resonance spectroscopy.[43] Note that numerous earlier models relied only on different combinations of X-ray diffraction, elemental composition and chemical reactivity of GO. Short-contact-time experiments suggested an existence of hydrogen bonds connecting parallel planes.[42] Hydrogen bonds between planes can made exfoliation more complicated. One can anticipate competitive hydrogen bonds of GO with the polar solvent in which exfoliation is conducted. The model of Lerf and Klinowski[43] constitutes a significant advance in the structure determination of GO. According to their experiments, water molecules are able to penetrate between the platelets and remain there for significant periods of time. Likely, water molecules form hydrogen bonds with polar surface groups. Tertiary alcohols and 1,2-ethers (epoxides) prevail at the surface of GO.[1, 9] This model is currently dominant in the community, although slight modifications and adjustments are routinely proposed. For instance, it is hypothesized that five and six-membered lactols exist on the periphery of the platelets, while esters of tertiary alcohols are formed on the graphitic surface. Cai and coworkers reported successful isotopic labeling of GO.[44] This offers a new avenue for structure determination exploiting spectroscopic techniques.

Here, we report highly accurate ab initio calculations on the energetics of graphite/graphene oxidation, and GO chemical, geometric and electronic structure. We employ the second-order Møller-Plesset perturbation theory, which is more rigorous and consistent than density functional theory, and which can be applied in a straightforward manner to finite systems. Therefore, the ab initio calculations are performed with graphene quantum dots (GQD). We find that formation of hydroxyl, carboxyl and epoxide groups is thermodynamically favorable both at the center and near the edge of the GQD. The edge location is preferred. The carbon-oxygen bond length between the GQD and the functional groups depends on the group position in the sheet. All oxygen containing groups perturb GQD planarity. We observe an unexpectedly strong reinforcement of chemical bonding due to electronic polarization of the GQD. This fundamental finding is likely to apply to other nanoscale systems. The reported results foster understanding of physical and chemical properties of GOs at the atomistic level of detail. By establishing the location and stability of oxygen containing groups on GO surface, the reported results are particularly useful for deciphering structure-determination experiments, e.g.,

X-ray. The data can be applied directly to predict GO electronic properties, required for various nanoscale devices.

**Methodology**

Electronic structures of the multiple investigated models of GQD, GO and auxiliary components were obtained by means of the second-order Møller-Plesset perturbation theory (MP2).[45] MP2 includes electron-correlation beyond the mean-field Hartree-Fock method and provides accurate electronic energy levels, band gaps, densities of states, stable-point geometries and binding energies. Consequently, MP2 is a more accurate and significantly more computationally demanding procedure, as compared to pure and hybrid density functional theory implementations.

The reported bond lengths, angles and partial charges correspond to thoroughly optimized structures following an energy gradient with a convergence threshold of $10^{-5}$ Hartree. The partial charges were derived using the GAMESS[46] output as shown elsewhere.[47] The split-valence triple-zeta 6-311+G* Pople-type basis set with polarization and diffuse functions was applied. The electronic energy convergence criterion at every self-consistent field step was set to $10^{-8}$ Hartree. The basis set superposition error was excluded from the binding energies using the counterpoise method.

Thermodynamic potentials of oxidation were calculated through frequency analysis and molecular partition functions using the established equations of statistical mechanics. It was important to use the electron-correlation method at this step, since the largest numerical discrepancies originate from neglect of electron correlation.

**Results and Discussion**

Upon oxidation, formation of multiple oxygen containing groups (Figure 1) should be anticipated. Depending on the valence requirements of a particular group and available connectivity of GQD, it can be located either at the surface or at the edge. We use a small GQD, $C_{32}H_{14}$, and consider three different oxidation sites: "center", "side" and "edge" (see the TOC image). We consider hydroxyl (-OH), carboxyl (-COOH) and epoxide (>O) groups (Figure 2) at these positions. As it was identified by means of electronic structure calculations at the MP2 level of theory, the carbonyl group (=O) is stable only at the "edge" position, while it readily re-arranges into an epoxide group when initially grafted at other positions. This oxygen containing group is stable and does not undermine stability of GQD. It can be used as a reference.

A classical and affordable approach was considered for the oxidation of GQD to obtain GO – treatment with potassium permanganate and sulfuric acid (Table 1). The products of this reaction are GO, potassium and manganese salts, and a certain amount of water, which is also a solvent in this reaction. In rare cases, a water molecule can be a reactant, although normally it is a product.

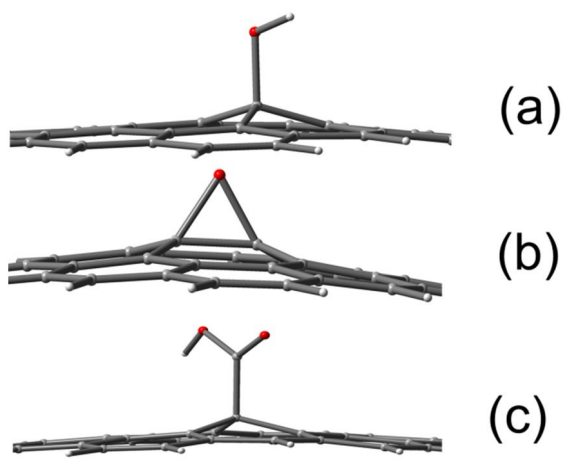

**Figure 1**. Optimized geometries of the oxidized GQD: (a) hydroxyl; (b) epoxide; (c) carboxyl groups grafted to one of the central atoms of GQD.

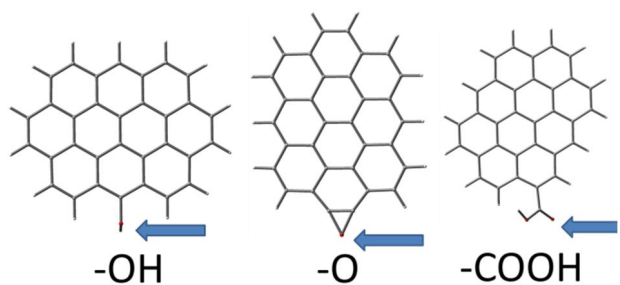

**Figure 2**. Optimized geometries of the oxidized GQD: (a) hydroxyl; (b) epoxide; (c) carboxyl groups grafted to one of the edge atoms of GQD.

The reactions where the carboxyl group is grafted are different, since they require an additional carbon atom. The source of this additional carbon atom can be manifold. In the edge position, the reaction $C_{32}H_{14} \rightarrow C_{32}H_{13}COOH$ can be compared with a conventional laboratory electrophilic substitution followed by oxidation: $C_6H_6 \rightarrow C_6H_5CH_3 \rightarrow C_6H_5COOH$ (Table 2). The stoichiometric coefficients of the reactions are presented in Tables 1-2. They are used for further energy analysis and comparison between different products.

**Table 1**. Reactants, products and numbers of their moles in the balanced chemical reactions, leading to oxidation of graphite (graphene layer). See Figures 1-2 for configurations of GO. The solvent (water) participates in all reactions. Depending on the products, water can be both a reactant (minus sign) and a product (plus sign)

| # | Reactants | | | Products | | | water |
|---|---|---|---|---|---|---|---|
|   | $C_{32}H_{14}$ | $H_2SO_4$ | $KMnO_4$ | GO | $K_2SO_4$ | $MnSO_4$ |  |
| 1 | 10 | 3 | 2 | 10 $C_{32}H_{14}OH$ | 1 | 2 | -2 |
| 2 | 5 | 3 | 2 | 5 $C_{32}H_{13}OH$ | 1 | 2 | +3 |
| 3 | 5 | 3 | 2 | 5 $C_{32}H_{14}O$ | 1 | 2 | +3 |
| 4 | 5 | 6 | 4 | 5 $C_{32}H_{13}O$ | 2 | 4 | +11 |
| 5 | 10 | 9 | 6 | 10 $C_{32}H_{12}O$ | 3 | 6 | +14 |
| 6 | 33 | 33 | 22 | 32 $C_{32}H_{14}COOH$ | 11 | 22 | +24 |
| 7 | 165 | 213 | 142 | 160 $C_{32}H_{13}COOH$ | 71 | 142 | +248 |

**Table 2**. Reactants, products and numbers of their moles in the balanced chemical reactions, leading to oxidation of benzene

| # | Reactants | | | Products | | | water |
|---|---|---|---|---|---|---|---|
|   | $C_6H_6$ | $H_2SO_4$ | $KMnO_4$ | $C_xH_yO_z$ | $K_2SO_4$ | $MnSO_4$ |  |
| 1 | 5 | 3 | 2 | 5 $C_6H_5OH$ | 1 | 2 | 3 |

| 2 | 10 | 9 | 6 | 10 $C_6H_5O$ | 3 | 6 | 14 |
| 3 | 5 | 6 | 4 | 5 $C_6H_4O$ | 2 | 4 | 11 |
| 4 | 7 | 9 | 6 | 6 $C_6H_5COOH$ | 3 | 6 | 12 |

If OX is grafted to the carbon atom at the center of GR, aromaticity in the corresponding six-membered ring is perturbed, since an additional electron is required to form a new covalent bond, C-O or C-C. Consequently, GQD deforms significantly to respond to a new chemical environment and to maintain stability under thermal motion. This is the reason why GO cannot be reduced precisely to pristine graphene, which constitutes a serious synthetic challenge.

Oxidation of the "edge" positions is very favorable compared to the "center" and "side" positions (Figure 3). The differences between the "center" and "edge" positions are small, although the atomistic environments are somewhat different. The "side" position is located next to the edge, in a ring with the passivating hydrogen atoms (TOC). Note that the –COOH group contains two oxygen atoms, while the other OX groups contain a single oxygen atom. Oxidation is particularly favorable at the "edge" position, since carbon-oxygen bonds are stronger than carbon-hydrogen bonds.

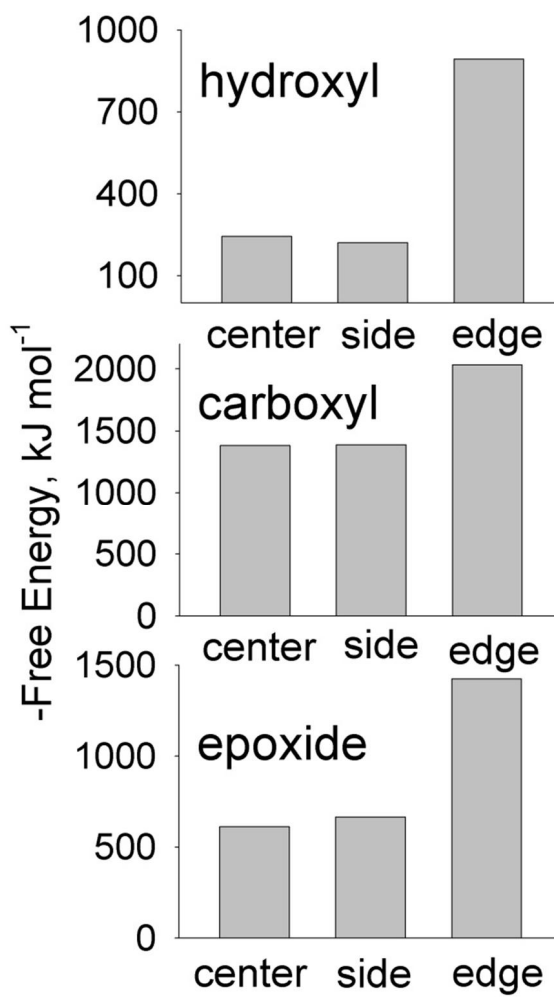

**Figure 3**. Free energy of oxidation of the $C_{32}H_{14}$ GQD at 300 K and 1 bar depending on the oxygen containing group. Free energies were multiplied by -1, so that a larger number corresponded to a more favorable reaction.

According to Figure 4, oxidation is driven by an enthalpy change, rather than by an entropic contribution, $-T \times S$, where T=300 K. If OX is formed at the edge of GQD, then entropy constitutes two or less per cents of the Gibbs free energy. However, if OX is located at the surface, then entropic contribution rises up to 12%. Entropy increases upon full oxidation (i.e. burning), but it decreases upon partial oxidation, such as that considered in the present work. In present, the entropy change reflects geometry alterations in GQD due to oxidation. It is somewhat larger at the "center" position than at the "side" position. Figures 5 and 6 analyze geometry perturbations brought by the oxidation, rationalizing correlations of structure and Gibbs energy.

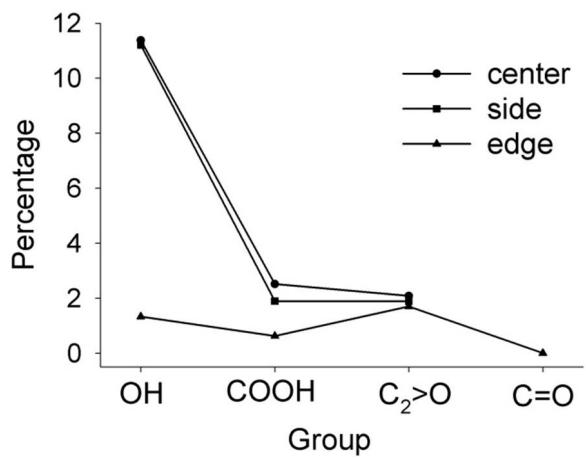

**Figure 4**. Percent contribution of the entropic factor into the total oxidation reaction free energy depending on the position of the oxygen containing group within GQD.

**Table 3**. Thermodynamic potentials for the oxidation reaction of benzene at standard conditions depending on the final product. The oxidation of benzene provides comparison with GO

| Reaction Product | Enthalpy, kJ mol$^{-1}$ | Entropy, J mol$^{-1}$ K$^{-1}$ | Free Energy, kJ mol$^{-1}$ |
|---|---|---|---|
| $C_6H_5OH$ | -491 | -46.7 | -477 |
| $C_6H_5O$ | -767 | +25.9 | -774 |
| $C_6H_4O$ | -1066 | +78.3 | -1089 |
| $C_6H_5COOH$ | -1551 | -41.3 | -1539 |

Figure 5 depicts bond lengths between the selected carbon atom of GQD and the carbon (-COOH) or oxygen ($C_2>O$, C-OH, C=O) atom of OX. These bonds are polar covalent. Figure 6 depicts C-C-C-C dihedrals in GQD after an oxidation. These dihedrals constitute an important measure of planarity. Analyzed together, Figures 5-6 provide information on the influence of the oxygen containing group on the GQD geometry. The carbon-carbon bond length in benzene is 0.140 nm. This value corresponds to an optimized geometry with thermal expansion and fluctuations excluded. The carbon-carbon bond in GQD is longer, 0.141 nm. The same length is observed in the case of carbon nanotubes. This difference is insignificant. It is in concordance with stability of benzene and other, more sophisticated representatives of carbonaceous materials. The bond lengths depend on the position within GQD: "center", "side" or "edge". Compare the distances with the carbon-oxygen bond in ethylene oxide, 0.141 nm, and carbon-oxygen bond in methanol, 0.143 nm. The "edge" position corresponds to the shortest bond. This also means that the "edge" position of oxygen is most energetically favorable in GO. The epoxide group represents an exception, although the difference in its case between various positions is very small. The epoxide group at the "edge" position substitutes two hydrogen atoms, while it does not substitute any atoms at other positions.

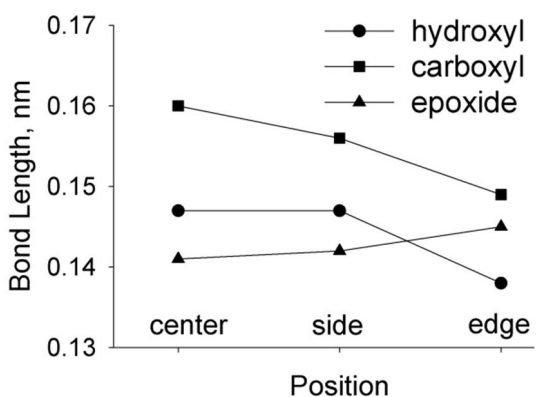

**Figure 5**. Covalent bond length between $C_{32}H_{14}$ GQD and QX (hydroxyl, carbonyl, epoxide). Compare the depicted data with the double bond length in the case of carbonyl group at the edge position, 0.123 nm.

The OX groups attached to one of the central atoms undermine planarity of GQD (Figure 6). The most drastic change is observed in the case of the -COOH group, ca. 30 degrees. The effect of the -OH group is smallest, although it also exceeds 20 degrees. Perturbation of the GQD planarity does not explain why entropy is more important in the formation of –OH than –COOH (Figure 4). This observation will be explained by the partial charge distribution below.

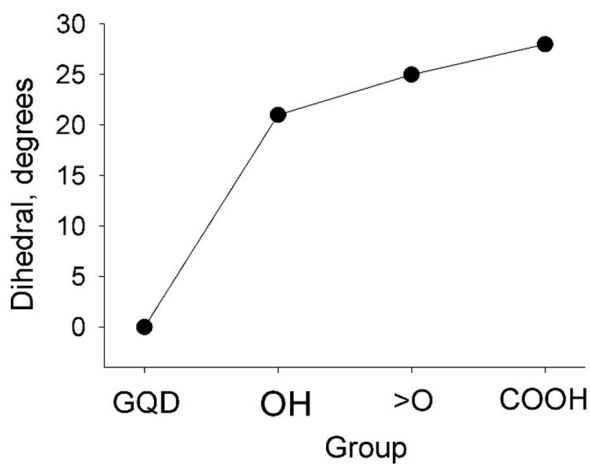

**Figure 6**. Averaged C-C-C-C dihedral angles of the GQD carbon atoms in the vicinity of OX (the "center" position).

Partial charges and dipole moments depend significantly on the position of OX. The results for the "center" and "side" positions are similar, while the results for the "edge" position are significantly different: Compare positive charges on GQD when the –OH group is attached to the central carbon atom and to the rim carbon atom. Interestingly, the largest difference is observed in the case of the –COOH group. Dipole moment increases from 1.2 D, when –COOH is at the "side" position, to 6.5 D when this group is at the 'edge' position. In turn, the position of the epoxide group does not exhibit any influence on the electronic density distribution.

The largest positive charges are induced on those carbon atoms, which are located in the vicinity of the –OH group, while the –COOH group polarizes an electronic density of the GQD plane quite modestly. This behavior is likely responsible for a higher entropic contribution into the formation of –OH, as compared to the formation of –COOH.

**Table 4**. Partial electrostatic charges in the vicinity of OX and dipole moments of the corresponding GO species

| Group/Position | Center | Side | Edge |
|---|---|---|---|
| Charge, e | | | |
| -OH | 0.39 | 0.41 | 0.14 |
| <O | 0.31 | 0.29 | 0.31 |
| -COOH | 0.15 | 0.16 | -0.02 |
| -C=O | unstable | unstable | 0.53 |
| Dipole moment, D | | | |
| -OH | 1.6 | 1.3 | 1.5 |
| <O | 2.2 | 2.2 | 2.2 |
| -COOH | 4.3 | 1.2 | 6.5 |
| -C=O | unstable | unstable | 5.3 |

Bond energy is one of the most interesting properties reported in the context of GO. Bond energy is a quantity of energy, which has to be absorbed upon separation of GQD and OX. Chemical reduction is applied widely to obtain reduced GO from the exfoliated graphite.

Reduction means removal of most OX. The strengths of polar covalent bonds between GQD and OX help to understand reduction products and their relative yields. Note that reduced GO is not structurally identical to GQD.

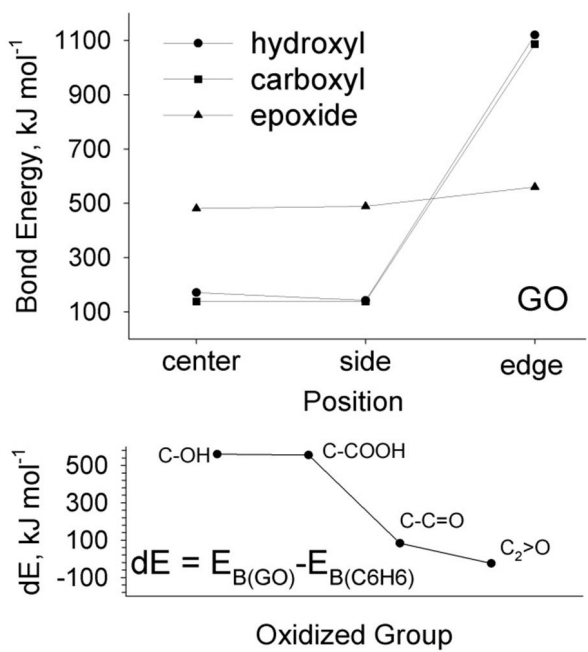

**Figure 7**. (Top) Energies of bonds between the carbon atom of GQD and the relevant atom of OX (hydroxyl, carboxyl, epoxide) depending on the position. (Bottom) Difference between the energies of certain bonds (see legend) in benzene and GO (the "edge" position). It is remarkable that the difference in the bond strength is very significant.

The reported bond energies depend not only on the covalent interaction between the rim carbon atom of GQD and OX, but also on an electrostatic attraction between point charges of OX and induced electrostatic charges on GQD. This type of bond strengthening can be regarded as a new type of chemical bond, since the corresponding energy (Figure 7) is much larger than that of C-OX bond alone. The comparison made with benzene reveals differences of above 500 kJ mol$^{-1}$! These energy supplements are non-covalent in nature and should be ascribed to an electrostatic contribution due to an electronic polarization of GR. The largest contributions were recorded in the cases of –COOH and –OH, whereas carbonyl and, especially, epoxide groups induce more modest effects. The unusually strong interaction between the oxygen containing groups and GQD can affect photo-induced electron transfer, electron-vibrational relaxation and related processes.[34]

The binding of OX at the "edge" positions is much stronger as compared to the "side" and "center" positions. Chemical bonding of OX to the central carbon atoms of GQD occurs by interaction with aromatic electrons and is thermodynamically permitted. Nevertheless, this bonding costs GQD significant energy. The resulting GO is thermally stable, but susceptible to reduction agents. In turn, reduction of the "edge" positions is very challenging, according to our present analysis. The difference between the "side" and "center" positions is nearly indistinguishable. The only exception is the –OH group, where the difference amounts to 29 kJ mol$^{-1}$ in favor of the "center" position. The simplest explanation of this fact is that the center of GQD is more electron rich and the side of GQD is electron poorer due to the terminating hydrogen atoms. The oxygen atom of -OH, which is most electronegative in the considered atomic ensemble, attracts electron density. Therefore, it prefers a more electron rich location. This effect, however, constitutes just 17% of the total binding energy. In the case of the –COOH group, the oxygen atom attracts electron density from the carboxyl carbon atom, and the "center" vs "side" effect is not observed at all.

The binding energy of GQD with the epoxide group is less dependent on its position within GQD. Two factors are responsible for this. First, the epoxide group creates two bonds with GQD, whereas all other groups create a single bond. The carbonyl group at the "edge" position creates a double bond. One may consider dividing the corresponding epoxide group energy by two for a more direct comparison with other groups. Second, the epoxide-only GO maintains a singlet electronic configuration, while the –OH and –COOH groups engender doublet configurations. Adding another doublet group in the vicinity of the first doublet group will likely stabilize both of them, to a certain extent. Note that calculation of binding energy in the case of two and more grafted OX is technically challenging, since the non-covalent interaction between the neighboring OX should be excluded.

**Concluding Remarks**

We have investigated interaction of several oxygen containing groups (hydroxyl, carboxyl, epoxide) with the surface of a GQD as a function of the group position (center, side, edge), using high level ab initio electronic structure methodology. We conclude that formation of all of these groups is thermodynamically permitted when a combination of strong oxidants, such as $KMnO_4$ and $H_2SO_4$, is employed. The edge is energetically more favorable than the side and center positions. The side and center energetics are similar, with minor differences attributable to a larger amount of electron density available at the GQD center. Attachment of the hydroxyl, carboxyl and epoxide groups to the GQD perturbs the planar GQD structure. The bond lengths between the GQD and the OX groups depend on the OX position within the sheet. Comparison between GO and benzene derivatives reveals an interesting and important phenomenon. The bond between GQD and OX is much stronger than the bond between benzene and the same OX. The effect can be explained by an additional electrostatic contribution due to an electronic polarization of GQD. This type of bonding reinforcement can be generic for many nanoscale materials and deserves further investigation.

Our findings advance the atomistic understanding of the chemical, geometric and electronic structure of graphene oxides, guiding both fundamental experiments and nano-technological applications. Experiments, such as X-ray spectroscopy, aimed at determination of GO structure are hard to interpret in a unique manner due to a broad variety of possible GO isomers. The reported thermodynamic data predict stability of various isomers, assisting in the interpretation. The analysis is performed for typical oxidation conditions in the presence of $H_2SO_4$ and $KMnO_4$. It is straightforward to use data for GR oxidation under alternative conditions, as well as for GO reduction and related chemical reactions. Knowledge of the stability and location of various oxygen containing groups on GO surface enables one to predict performance of sensing, optoelectronic, photovoltaic, catalytic, drug delivery. and other nanoscale devices.

**Acknowledgments.** V.V.C. is CAPES fellow (Brazil). OVP acknowledges financial support of the U.S. National Science Foundation, grant CHE-1300118.

**Contact Information.** E-mail for correspondence: vvchaban@gmail.com. Tel: +55 12 3309-9573; Fax: +55 12 3921-8857.

**REFERENCES**

1. D. R. Dreyer, S. Park, C. W. Bielawski and R. S. Ruoff, *Chem Soc Rev*, 2010, 39, 228-240.
2. Y. Hong, L. Li, X. C. Zeng and J. C. Zhang, *Nanoscale*, 2015, 7, 6286-6294.
3. M. H. Wu, X. C. Zeng and P. Jena, *Journal of Physical Chemistry Letters*, 2013, 4, 2482-2488.
4. N. Prabhakar, T. Nareoja, E. von Haartman, D. Sen Karaman, S. A. Burikov, T. A. Dolenko, T. Deguchi, V. Mamaeva, P. E. Hanninen, Vlasov, II, O. A. Shenderova and J. M. Rosenholm, *Nanoscale*, 2015, 7, 10410-10420.
5. J. Azevedo, L. Fillaud, C. Bourdillon, J. M. Noel, F. Kanoufi, B. Jousselme, V. Derycke, S. Campidelli and R. Cornut, *J Am Chem Soc*, 2014, 136, 4833-4836.
6. T. F. Liu, D. Kim, H. W. Han, A. B. Yusoff and J. Jang, *Nanoscale*, 2015, 7, 10708-10718.
7. I. P. Murray, S. J. Lou, L. J. Cote, S. Loser, C. J. Kadleck, T. Xu, J. M. Szarko, B. S. Rolczynski, J. E. Johns, J. X. Huang, L. P. Yu, L. X. Chen, T. J. Marks and M. C. Hersam, *Journal of Physical Chemistry Letters*, 2011, 2, 3006-3012.
8. C. L. Lee and I. D. Kim, *Nanoscale*, 2015, 7, 10362-10367.
9. D. R. Dreyer, A. D. Todd and C. W. Bielawski, *Chem Soc Rev*, 2014, 43, 5288-5301.
10. L. Gao, Q. Li, R. Q. Li, L. R. Yan, Y. Zhou, K. P. Chena and H. X. Shi, *Nanoscale*, 2015, 7, 10903-10907.
11. K. K. Mao, L. Li, W. H. Zhang, Y. Pei, X. C. Zeng, X. J. Wu and J. L. Yang, *Scientific Reports*, 2014, 4, 5441.
12. J. Zhang, Q. Wang, L. H. Wang, X. A. Li and W. Huang, *Nanoscale*, 2015, 7, 10391-10397.
13. S. Y. Lee and S. J. Park, *Carbon Lett*, 2012, 13, 73-87.
14. G. Jalani and M. Cerruti, *Nanoscale*, 2015, 7, 9990-9997.
15. P. A. Bharad, K. Sivaranjani and C. S. Gopinath, *Nanoscale*, 2015, 7, 11206-11215.
16. K. Yao, M. Manjare, C. A. Barrett, B. Yang, T. T. Salguero and Y. P. Zhao, *Journal of Physical Chemistry Letters*, 2012, 3, 2204-2208.
17. M. R. Karim, K. Hatakeyama, T. Matsui, H. Takehira, T. Taniguchi, M. Koinuma, Y. Matsumoto, T. Akutagawa, T. Nakamura, S. Noro, T. Yamada, H. Kitagawa and S. Hayami, *J Am Chem Soc*, 2013, 135, 8097-8100.
18. J. Z. Xu, Y. Y. Liang, G. J. Zhong, H. L. Li, C. Chen, L. B. Li and Z. M. Li, *Journal of Physical Chemistry Letters*, 2012, 3, 530-535.
19. J. T. Robinson, S. M. Tabakman, Y. Y. Liang, H. L. Wang, H. S. Casalongue, D. Vinh and H. J. Dai, *J Am Chem Soc*, 2011, 133, 6825-6831.
20. J. Y. Luo, L. J. Cote, V. C. Tung, A. T. L. Tan, P. E. Goins, J. S. Wu and J. X. Huang, *J Am Chem Soc*, 2010, 132, 17667-17669.
21. D. A. Sokolov, K. R. Shepperd and T. M. Orlando, *Journal of Physical Chemistry Letters*, 2010, 1, 2633-2636.
22. E. Tylianakis, G. M. Psofogiannakis and G. E. Froudakis, *Journal of Physical Chemistry Letters*, 2010, 1, 2459-2464.
23. J. Kim, L. J. Cote, F. Kim, W. Yuan, K. R. Shull and J. X. Huang, *J Am Chem Soc*, 2010, 132, 8180-8186.
24. S. Wang, L. A. L. Tang, Q. L. Bao, M. Lin, S. Z. Deng, B. M. Goh and K. P. Loh, *J Am Chem Soc*, 2009, 131, 16832-16837.
25. Y. X. Li, C. Y. Sun, C. J. Yu, C. X. Wang, Y. J. Liu and Y. B. Song, *Adv Mater Res-Switz*, 2012, 476-478, 1488-1495.
26. O. Penkov, H. J. Kim, H. J. Kim and D. E. Kim, *Int J Precis Eng Man*, 2014, 15, 577-585.
27. S. Mao, K. H. Yu, S. M. Cui, Z. Bo, G. H. Lu and J. H. Chen, *Nanoscale*, 2011, 3, 2849-2853.
28. L. L. Zhang, C. L. Jiang and Z. P. Zhang, *Nanoscale*, 2013, 5, 3773-3779.
29. J. Hong, K. Char and B. S. Kim, *Journal of Physical Chemistry Letters*, 2010, 1, 3442-3445.
30. O. V. Prezhdo, *Surf Sci*, 2011, 605, 1607-1610.
31. B. R. Burg and D. Poulikakos, *J Mater Res*, 2011, 26, 1561-1571.
32. R. Long, N. J. English and O. V. Prezhdo, *J Am Chem Soc*, 2012, 134, 14238-14248.
33. M. J. Allen, V. C. Tung and R. B. Kaner, *Chem Rev*, 2010, 110, 132-145.
34. T. R. Nelson and O. V. Prezhdo, *J Am Chem Soc*, 2013, 135, 3702-3710.

35. L. Cardenas, J. MacLeod, J. Lipton-Duffin, D. G. Seifu, F. Popescu, M. Siaj, D. Mantovani and F. Rosei, *Nanoscale*, 2014, 6, 8664-8670.
36. H. Y. Jeong, J. Y. Kim, J. W. Kim, J. O. Hwang, J. E. Kim, J. Y. Lee, T. H. Yoon, B. J. Cho, S. O. Kim, R. S. Ruoff and S. Y. Choi, *Nano Lett*, 2010, 10, 4381-4386.
37. H. Joshi, K. N. Sharma, A. K. Sharma and A. K. Singh, *Nanoscale*, 2014, 6, 4588-4597.
38. T. W. Kim, Y. Gao, O. Acton, H. L. Yip, H. Ma, H. Z. Chen and A. K. Y. Jen, *Appl Phys Lett*, 2010, 97.
39. F. Perrozzi, S. Prezioso and L. Ottaviano, *J Phys-Condens Mat*, 2015, 27.
40. Y. W. Zhu, S. Murali, W. W. Cai, X. S. Li, J. W. Suk, J. R. Potts and R. S. Ruoff, *Adv Mater*, 2010, 22, 3906-3924.
41. N. R. Wilson, P. A. Pandey, R. Beanland, R. J. Young, I. A. Kinloch, L. Gong, Z. Liu, K. Suenaga, J. P. Rourke, S. J. York and J. Sloan, *Acs Nano*, 2009, 3, 2547-2556.
42. M. Mermoux, Y. Chabre and A. Rousseau, *Carbon*, 1991, 29, 469-474.
43. A. Lerf, H. Y. He, M. Forster and J. Klinowski, *J Phys Chem B*, 1998, 102, 4477-4482.
44. W. W. Cai, R. D. Piner, F. J. Stadermann, S. Park, M. A. Shaibat, Y. Ishii, D. X. Yang, A. Velamakanni, S. J. An, M. Stoller, J. H. An, D. M. Chen and R. S. Ruoff, *Science*, 2008, 321, 1815-1817.
45. M. Head-Gordon, J. A. Pople and M. J. Frisch, *Chem Phys Lett*, 1988, 153, 503-506.
46. M. W. Schmidt, K. K. Baldridge, J. A. Boatz, S. T. Elbert, M. S. Gordon, J. H. Jensen, S. Koseki, N. Matsunaga, K. A. Nguyen, S. J. Su, T. L. Windus, M. Dupuis and J. A. Montgomery, *J Comput Chem*, 1993, 14, 1347-1363.
47. F. Y. Dupradeau, A. Pigache, T. Zaffran, C. Savineau, R. Lelong, N. Grivel, D. Lelong, W. Rosanski and P. Cieplak, *Phys Chem Chem Phys*, 2010, 12, 7821-7839.